\documentclass[twocolumn, secnumarabic, amssymb, noshowpacs,nobibnotes, aps, showpacs,prb]{revtex4-1}
\usepackage{graphics}
\usepackage[dvips]{graphicx}
\usepackage{dcolumn}
\usepackage{amsmath}

\begin{document}

\title{New approach to study light-emission of periodic structures. Unveiling novel surface-states effects}
\author{Pedro Pereyra}
\address{F\'{i}sica Te\'{o}rica y Materia Condensada, UAM-Azcapotzalco,  C.P. 02200, M\'{e}xico D. F., M\'{e}xico }
\date{\today}

\begin{abstract}
An accurate approach to calculate the optical response of periodic structures is proposed. Using the genuine superlattice eigenfunctions and energy eigenvalues, the eigenfunctions parity symmetries, the subband symmetries and the detached surface energy levels, we report new optical-transition selection rules and explicit optical-response calculations. Observed transitions that were considered forbidden, become allowed and interesting optical-spectra effects emerge as fingerprints of intra-subband and surface states. The unexplained groups and isolated narrow peaks observed in high resolution blue-laser spectra, by Nakamura et al., are now fully explained and faithfully reproduced.

\end{abstract}
\pacs{03.65.Ge, 42.50.-p, 42.50.Ct, 42.62.Fi, 68.65.Ac, 73.20.-r, 78.30.Fs, 78.55.-m, 78.66.Fd, 78.67.Pt, 85.60.-q}

\maketitle


Although the fascinating phenomenon of light emission has been studied for more than a century, the main problem in calculating optical responses of semiconductor periodic structures using, for example, the golden rule
\begin{eqnarray}
|\langle\psi_{\rm f} |H_{\rm int}|\psi_{\rm i}\rangle |^2/[E_{\rm f}-E_{\rm i}+\hbar \omega)^2+\Gamma_{\rm i}^2]
\end{eqnarray}
where $H_{\rm int}$ describes the light-matter interaction and $\omega$ the emitted photon frequency, has been the lack of explicit knowledge of the initial and final states $|\psi_{i}\rangle$ and $|\psi_{f}\rangle $ and of the corresponding energies $E_{i}$ and $E_{f}$. \cite{EsakiLesHuches,Bastard} In the standard
approaches (SAs) to periodic systems, based on models and theorems for infinite periodic systems,\cite{Dingle1974,Mukherji1975,Dingle1975ASSP,SaiHalasz1978,LuoFurdyna1990,ChangSchulman,
Yang,Helm93,HaugKoch,Band,Virgilio} the energy levels become continuous  bands or
subbands (SBs).  This return to a quasi-continuous-energy description is responsible, for example, for the theoretical inability to explain  high resolution photolumnescence-spectra features, like those observed by Nakamura et al.\cite{NakamuraBook} and for approximate selection rules\cite{Dingle1975ASSP,Helm1991,LuoFurdyna1990} that lead to affirm that some observed transitions are ``forbidden ones". \cite{Molenkamp1988,SandersChang,Masselink,Reynolds1988,Fu,Zhu1995}

We show here that using the eigenvalues $E_{\mu \nu}$ and aigenfunctions $\Psi_{\mu \nu}(z)$, rigorously obtained in the theory of finite periodic systems,\cite{Pereyra2005} and the eigenfunctions' parity symmetries recently derived,\cite{Symmetries} we not only replace the continuous subbands description of the standard approach by the most accurate discrete subbands description, we also unveil the surface energy levels (see figure 1), responsible for the, so far, unexplained optical-spectra effects observed in high resolution experiments. At the same time we recover a truly quantum description of optical emissions in periodic structures.

\begin{figure}
\begin{center}
\includegraphics[width=230pt]{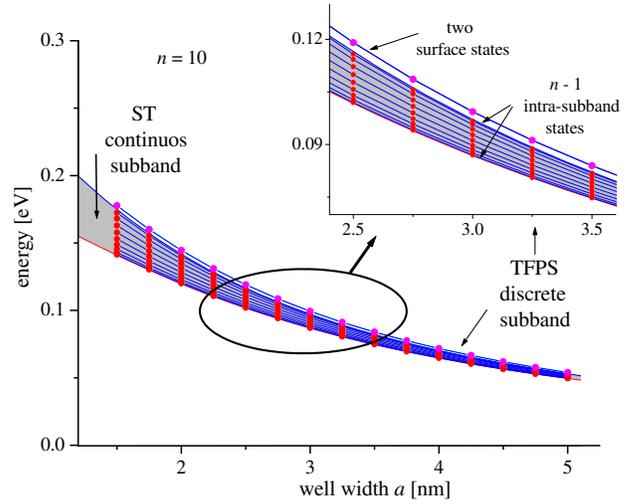}
\caption{Continuous and discrete first subband of the conduction band, as functions of the well width $a$, obtained for a Koronig-Penney-like periodic potential in the standard approach and in the TFPS, respectively. $n$-1 of the $n$+1 eigenvalues are true intra-subband energy levels and the remaining two are the energy levels of the surface states.}
\end{center}\label{Fig1}
\end{figure}
We will show that the detachment of the surface energy levels, apparent in figure 1,  is responsible for the groups of peaks observed by Nakamura et al. We will present  new selection rules, based on the eigenfunctions' symmetries and strongly dependent on the quatum numbers $\mu$ and $\nu$, the parity of $n$ and on the surface states. The ``forbidden transitions" will become allowed. We will report here two types of selection rules.  The first one, based on the eigenfunction's parity symmetry, will reduce the number of evaluations from $N\simeq(n+1)^2n_cn_v$ to $N/2$, for a SL with $n$ unit cells, $n_c$ SBs in the conduction band (CB) and $n_v$  SBs in the valence band (VB). The second rule, based on the subband symmetry. This rule will reduce the number of evaluations to $\simeq n n_cn_v/2$.

\begin{figure}
\begin{center}
\includegraphics [width=240pt]{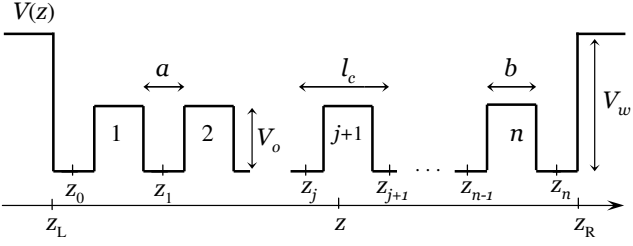}
\caption{Potential profile in the CB edge and parameters of a type I quasi-bounded superlattice. The wave function in Eq. (\ref{e.15}) applies to a point $z$ of the $j\!+\!1$ cell, with $0\leq j \leq (n-1)$.} \label{Fig2}
\end{center}
\end{figure}

For simplicity we will refer here to type I SLs. The generalization  is direct. It was shown in Ref.[\onlinecite{Pereyra2005}] that the eigenvalues, for SL bounded by cladding layers like in figure \ref{Fig2}, can be obtained  from
\begin{eqnarray}\label{EqEigenv}
\!\!\mathfrak{R}{\rm e}\left(\alpha_ne^{ika}\right)\!-\!\frac{k^{2}\!-\!q_{w}^{2}}{2q_{w}k}\mathfrak{I}{\rm m}\left(\alpha_ne^{ika}\right)\!-\!\frac{k^{2}\!+\!q_{w}^{2}}{2q_{w}k}\mathfrak{I}{\rm m}\beta_{n}\!=\!0,\hspace{0.2in}
\end{eqnarray}
where $q_w$ and $k$ are the wave numbers at the left (right) and right (left) of the discontinuity point $z_L$ ($z_R$),
$\alpha_n=U_n-\alpha^*U_{n-1}$  and $\beta_{n}=\beta U_{n-1}$ the $n$-cell transfer matrix elements, and $U_n$ the Chebychev polynomial of the second kind evaluated at the real part of the matrix element $M_{1,1}$ of the unit-cell transfer matrix
\begin{eqnarray}
M(z_{i+1},z_{i})=\left( \begin{array}{cc} \alpha & \beta \cr \beta^* & \alpha^*  \end{array}\right).
\end{eqnarray}
The eigenfunctions,  are obtained from
\begin{eqnarray}\label{EquEigFunc1}
\Psi _{\mu ,\nu }^{qb}(z)=\Psi ^{qb}(z,E_{\mu ,\nu }),
\end{eqnarray}
where
\begin{eqnarray}
\Psi^{qb}(z,E) &\!\!=\!\!&\!\frac{a_o}{2k}\Bigl[\Bigl((\alpha_{p}\!+\!\gamma
_{p})\alpha
_{j}\!+\!(\beta_{p}\!+\!\delta_{p})\beta_{j}^{\ast}\Bigr)e^{\!i k a/2}(k\!-\!iq_{w}
)\Bigr. \nonumber \\ &\!\!+\!\!&\Bigl.\Bigl((\alpha
_{p}\!+\!\gamma_{p})\beta_{j}\!+\!(\beta_{p}\!+\!\delta _{p})\alpha _{j}^{\ast
}\Bigr) e^{\!-i k a/2}(k\!+\!iq_{w})\Bigr].\nonumber \\
 \label{e.15}
\end{eqnarray}
Here $a_o$ is a normalization constant
and $z$ any point in the $j+1$ cell, i.e. any point between $z_j$ and $z_{j+1}$, with $0\leq j \leq (n-1)$. $\alpha_j$,
$\beta_j$,...  are the $j$-cells transfer-matrix elements and $\alpha_p$,
$\beta_p$,...the matrix elements of $M_p(z,z_j)$ that connects the state vectors $\Phi(z_j)$ and $\Phi(z)$, for $z_j\leq z \leq z_{j+1}$. The super-index $q$ refers to quasi-bound superlattice and $b=c,v$ refers to conduction and valence band. The super-index $q$ and the band index will be written only if they are necessary.

To evaluate the SL optical response, especifically the photoluminescence (PL) for specific systems, we will consider the golden rule
\begin{eqnarray}
\chi^r_{\small PL}\!=\! \sum_{\nu,\nu'\!,\mu,\mu'}f_{eh}\frac{\displaystyle \Bigl{|}\int dz
[\Psi^{v}_{\mu',\nu'}(z)]^*\frac{\partial}{\partial
z}\Psi^{c}_{\mu,\nu}(z)\Bigr{|}^{2}}{(\hbar \omega\!-\!E_{\mu,\nu}^{c}\!-\!E_g\!+\!E_{\mu',\nu'}^{v}\!+\!E_B)^{2}+\Gamma^{2}},\hspace{0.3in}
\label{susceptPL}
\end{eqnarray}
with energies measured from the corresponding band edges. Here $E_g$ is the gap energy,  $E_B$ the exciton binding energy, $\Gamma$ the level broadening energy and $f_{eh}$ the  occupation probability.

The parity symmetries, for eigenfunctions of quasi-bounded SLs,  are summarized as\cite{Symmetries}
\begin{eqnarray}\label{QBWFSym}
\Psi_{\mu,\nu}(z)\!=\!\Biggl\{\begin{array}{cc} (-1)^{\nu+1}\Psi_{\mu,\nu}(-z) & \text{for}\hspace{0.1in} n \hspace{0.1in}\text{odd}\cr  & \cr (-1)^{\nu+\mu}\Psi_{\mu,\nu}(-z) & \text{for}\hspace{0.1in} n \hspace{0.1in}\text{even}  \end{array}\Biggr..
\end{eqnarray}
These relations lead to the following symmetry selection rules (SSRs). For  $n$  even, we have:
\begin{widetext}
\begin{eqnarray}\label{selectrul1}
 \int dz
\Psi^{v}_{\mu',\nu'}(z)\frac{\partial}{\partial z}\Psi^{c}_{\mu,\nu}(z)\left\{ \begin{array}{llrl}
= 0 &\hspace{0.2in}{\rm when} \hspace{0.2in} P[\mu'+\nu']&=& P[\mu+\nu];\cr
\neq 0 &\hspace{0.2in}{\rm when}\hspace{0.2in} P[\mu'+\nu']&=& P[\mu+\nu+1].\end{array}\right.
\end{eqnarray}
Here $P[l]$ means parity of $l$. When $n$ is odd the SSRs are:
\begin{eqnarray}\label{selectrul2}
 \int dz
\Psi^{v}_{\mu',\nu'}(z)\frac{\partial}{\partial z}\Psi^{c}_{\mu,\nu}(z)\left\{ \begin{array}{llrl}
= 0 &\hspace{0.2in}{\rm when} \hspace{0.2in} P[\nu']&=& P[\nu];\cr
\neq 0 &\hspace{0.2in}{\rm when}\hspace{0.2in} P[\nu']&=& P[\nu+1].\end{array}\right.
\end{eqnarray}
\end{widetext}

Similar relations hold for IR transitions, with the additional restrictions $\mu \geq \mu'$ and, whenever $\mu=\mu'$, we must  also have $\nu > \nu'$, see Ref. [\onlinecite{SelecRules}]. These rules, as mentioned before, effectively reduce the number of possible transitions to $N/2$. Depending on the number of subbands, this can be still a large number. To reduce even more the number of matrix-elements evaluations,  we will introduce, some lines below, other rules related with the subband symmetry.

\begin{figure}
\includegraphics [width=270pt]{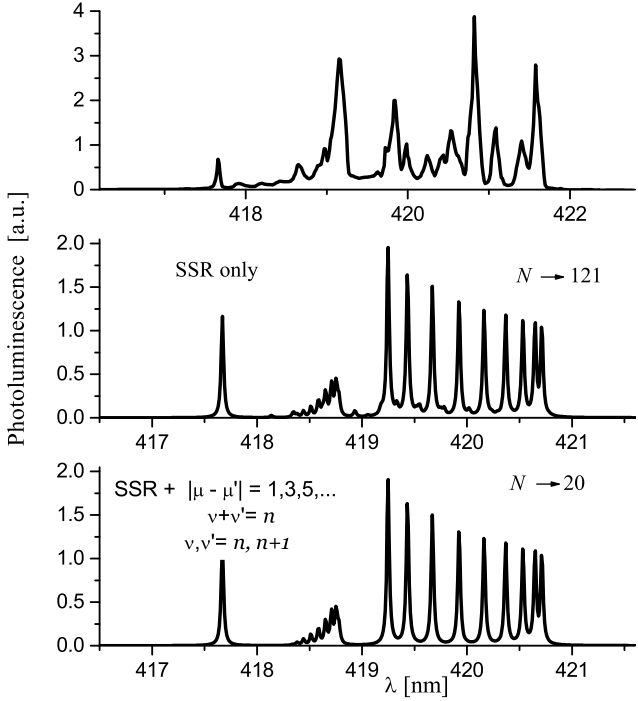}
\caption{Predicted and experimental PL spectra. Narrow and grouped peaks in the high resolution spectra measured by Nakamura et al.\cite{NakamuraBook,NakamuraPaper} (upper panel) for the blue emitting  $GaN\backslash(In_{0.2}Ga_{0.8}N\backslash
In_{0.05}Ga_{0.95}N)^{n}\backslash GaN$ superlattice with $n$=10, $a$=2.5nm and $b$=5nm, and our theoretical calculation (middle and lower panels). The experimental spectrum is reproduced with permission from [\onlinecite{NakamuraPaper}]. Copyright [1996], AIP Publishing LLC.}\label{Fig3}
\end{figure}

\begin{figure}
\includegraphics [width=220pt]{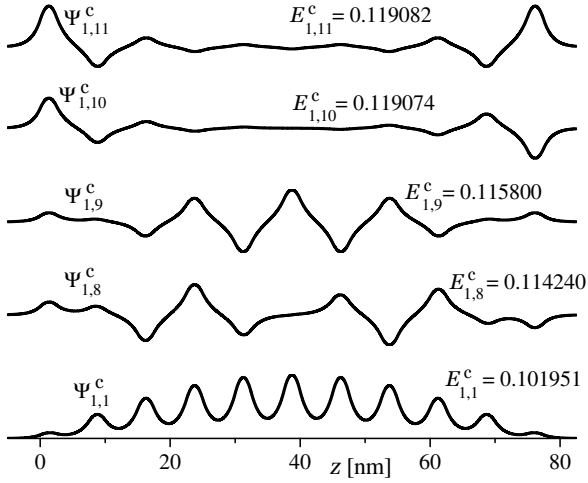}
\caption{Eigenfunctions and surface states. The eigenfunctions  $\Psi _{1 ,1 }^{c}(z)$,  $\Psi _{1 ,8 }^{c}(z)$ and $\Psi _{1 ,9 }^{c}(z)$, and the slightly detached surface states  $\Psi _{1 ,10 }^{c}(z)$, and  $\Psi _{1 ,11 }^{c}(z)$ in the first subband of the CB of the blue emitting $(In_{0.2}Ga_{0.8}N\backslash
In_{0.05}Ga_{0.95}N)^{10}\backslash In_{0.2}Ga_{0.8}N$ SL with  $a$=2.5nm and $b$=5nm, bounded by $GaN$ cladding layers.}\label{EigenfunctCBsb1}\label{Fig4}
\end{figure}

To test  our approach, we will consider two specific examples, with results obtained with highest experimental resolution that we could find in the literature. In Ref. [\onlinecite
{NakamuraBook}] the blue emitting SLs $(In_{x}Ga_{1-x}N\backslash
In_{y}Ga_{1-y}N)^{n}\backslash In_{x}Ga_{1-x}N$, bounded by $GaN$ and $AlGaN$ cladding layers, with $x$=0.2, $y$=0.05 and different values of $n$, have been extensively studied. Some results, show spectral features, with groups of narrow spectral widths and peak separations of the order of 0.2nm ($\sim$ 0.12meV), that could not be explained so far.

In the upper panel of figure 3, the PL spectrum,  first published in Ref. [\onlinecite{NakamuraPaper}], for a  SL with $n$=10 and using a monochromator  resolution\cite{NakamuraPaper} of  0.016nm ($\sim$ 0.01meV) is shown. Taking into account this SL parameters, and the appropriate electron and hole effective masses, we obtain the energy eigenvalues, the eigenfunctions and the PL spectrum plotted in the middle panel of  figure \ref{Fig3}. Some eigenfunctions $\Psi_{1,\nu}^c$, in the first SB of the CB, are plotted in figure \ref{EigenfunctCBsb1}. Notice that the eigenvalues $E_{1,10}^c$ and $E_{1,11}^c$, that correspond to the  surface states $\Psi_{1,10}^c$ and $\Psi_{1,11}^c$, are detached from the others energy levels in the SB.

To understand the structure of the optical response in figure  \ref{Fig3} let us distinguish, in each subband $\mu$ of the CB, the surface energy levels $\{s\mu\}$ from the remaining $n$-1 energy levels $\{g\mu\}$. Similarly, the energy levels  $\{s\mu'\}$ from the levels $\{g\mu'\}$, in the subband $\mu'$ of the VB. The transitions $g1 \rightarrow g2'$ are responsible for the group of peaks with larger wavelengths, between 419.24nm and 420.747nm, the transitions $g1 \rightarrow s2'$ and $s1 \rightarrow g2'$, for the group of peaks in the middle, and the transitions $s1 \rightarrow s2'$  for the isolated peak at the left.

This non-obvious resonant structure is a consequence of the presence of surface states, whose detachment determines the shift and the appearance of groups of peaks, as well as, of the isolated peak at a higher energy. It is clear that in order to observe this effect we need high-resolution experiments.

A rather general characteristic  of the PL and IR spectra, measured or calculated, is the small number of peaks, much smaller than the $N/2$. One reason is, of course, the low experimental resolution. From the explicit calculations, we found  out that, besides the parity symmetry, we have also the subband symmetry, glimpsed in Ref. [\onlinecite{Pereyra2005}], playing an important role in the relative values of the transition-matrix elements. In fact, when the surface levels detach, the matrix elements that fulfill the conditions
\begin{eqnarray}\label{LORDetached}
\langle {\mu',\nu'}|\frac{\partial}{\partial z}|{\mu,\nu}\rangle \hspace{0.1in} {\rm with}\hspace{0.1in}\begin{array}{rcl} |\mu-\mu'|\!&\!=\!&\!1,3,5,...\cr & {\rm and} &\cr \nu+\nu'\!&\!=\!&\!n \cr \nu=n,n+1&&\nu'=1,2,... \cr \nu'=n,n+1&&\nu=1,2,...,\end{array}
\end{eqnarray}
are leading order transitions. When the surface levels do not detach, the leading order transitions are
\begin{eqnarray}\label{LORUndetached}
\langle {\mu',\nu'}|\frac{\partial}{\partial z}|{\mu,\nu}\rangle \hspace{0.1in} {\rm where}\hspace{0.1in}\begin{array}{rcl} |\mu-\mu'|&=&1,3,5,...\cr & {\rm with} &\cr \nu+\nu'&=&n,n+2. \end{array}
\end{eqnarray}
\begin{figure}[t]
\includegraphics [width=220pt]{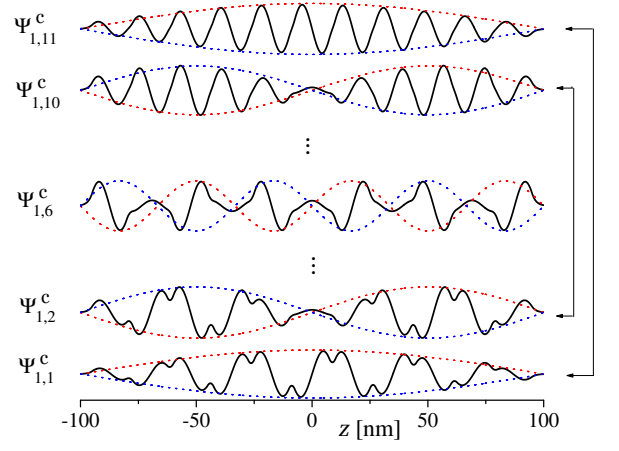}
\caption{The subband symmetry. Because of this symmetry the envelopes of the eigenfunctions with indices ($\mu,\nu$) and ($\mu,n$+2-$\nu$) are the same.  See for example $\Psi_{2,2}$ and $\Psi_{2,10}$. Here we plot the eigenfunctions $\Psi_{2,\nu}$ for a SL with $n$=10, thus with $\nu$=1, 2, ..., 11. }\label{Fig24}
\end{figure}
Because of the subband symmetry, the envelope curve of $\Psi_{\mu,\nu}$ is similar to that of $\Psi_{\mu,n+2-\nu}$, when the SSs do not detach, and similar to that of $\Psi_{\mu,n-\nu}$, when they detach. The eigenfunctions in figure \ref{Fig24}, correspond to a system where the SSs do not detach. In the blue emitting SL studied here the SSs detach, see figure \ref{Fig4}.

The leading order rules (LORs) reduce the number of matrix-elements evaluations. For a PL spectrum, the reduction is from $N/2\simeq(n+1)^2n_cn_v/2$ to $\simeq n n_c n_v$, in the first case, and to $(n+1)n_cn_v$, in the second.  To obtain the IR spectra we have, additionally, the condition $\mu'\leq\mu$.

If we apply the LORs for our example, with $n=10$, $n_c$=1 and $n_v$=2, the number of matrix-elements evaluations reduces from $(n+1)^2n_v/2$=121 to $nn_v$=20. The new spectrum, shown in the lower panel of figure \ref{Fig3}, contains essentially the same information as the one in the center. At the end, the number of transition matrix-elements that we calculate, using the SSRs and the LORs, is the same  as the number of peaks in the actual PL spectra.
\begin{figure}
\includegraphics [width=240pt]{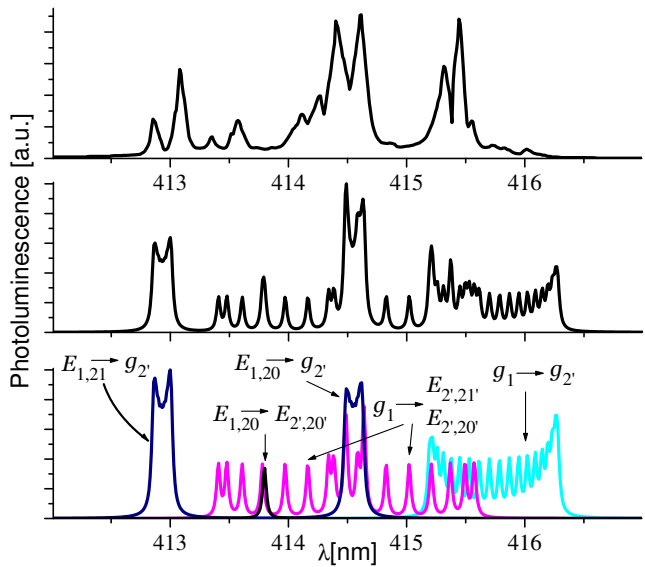}
\caption{Experimental\cite{NakamuraPaper2}  and theoretical PL spectra, upper and middle panels respectively, for a $GaInN$ SL similar to the one considered for figure \ref{Fig3}, but here with $n=20$ and asymmetric confining potential, thus larger SSs detachment. In the lower panel we plot separately, and indicate, the transitions that contribute to the PL in the middle panel. It is clear that without the detachment of the surface energy levels, $E_{1,21}$, $E_{1,20}$, $E_{2',20'}$ and $E_{2',21'}$, we would not account the experimental behavior and we will have a spectrum similar to that of the transitions $g1 \rightarrow g2'$, only. The experimental curve with permission of The Japan Society of Applied Physics, Copyright 1996.}\label{InGaNn20}
\end{figure}

In figure \ref{InGaNn20} we have the experimental and theoretical PL spectra for another sample in Yakamura's et al. book, first published in Ref. [\onlinecite{NakamuraPaper2}]. In this case the number of unit cells, $n$=20, implies 882 matrix-elements evaluations. In the upper panel we show the spectrum (c) of figure 11.10 in Reference [\onlinecite{NakamuraBook}].
To account for this result we had to take into account the confining potential asymmetry. Because of this asymmetry, the surface energy levels split and the surface states loos their parity symmetry. The other eigenvalues and eigenfunctions remain almost unchanged. Thus, the SSRs are still valid for the $g1 \rightarrow g2'$ transitions, but not for the $E_{1,21} \rightarrow g2'$ and $E_{1,20} \rightarrow g2'$ transitions. The surface states $\Psi_{1,20}$ and $\Psi_{1,21}$ in $s1$ become localized at the opposite sides of the superlattice. The same happens with $\Psi_{2',20'}$ and $\Psi_{2',21'}$. Even so, using the SSRs and the LORs we end up calculating 78 matrix-elements and with the spectrum in the panel at the middle of figure \ref{InGaNn20}. As shown in the lower panel of this figure, the transitions $E_{1,21} \rightarrow g2'$ and $E_{1,20} \rightarrow g2'$, lead to the most visible structures, referred to as "broaden emission lines" in [\onlinecite{NakamuraPaper2}]. In these graphs we do not show the transition $E_{1,21} \rightarrow E_{2',21'}$, which occurs at higher energy.

Interesting surface-states effects were unveiled and the eigenfunctions' parity and subband symmetries role, on the selection rules and leading order rules, were shown. High-accuracy PL experimental results, with features that could not be explained before,\cite{NakamuraBookp268} are now fully understood. The improved optical response theory opens up the possibility to enhance the optical techniques for specific applications. We expect that the relation between surface states, cladding layers' energy gap and the novel group structure and isolated peak in the PL spectra, will be further studied and experimentally confirmed.

The author acknowledges useful comments of H. P. Simanjuntak, A. Robledo-Martinez, J. Grabinsky and E. Ley-Koo.

\end{document}